\documentclass[twocolumn,showpacs,aps,prl,superscriptaddress]{revtex4}    
\usepackage{graphicx}
\usepackage{dcolumn}
\usepackage{amsmath}
\usepackage{epsfig}    
\usepackage{rotating}

\RequirePackage{xspace}

\usepackage{relsize}
\def\babar{\mbox{\slshape B\kern-0.1em{\smaller A}\kern-0.1em
    B\kern-0.1em{\smaller A\kern-0.2em R}}}

\def\epem       {\ensuremath{e^+e^-}\xspace}

\def\mumu       {\ensuremath{\mu^+\mu^-}\xspace}

\def\ellell     {\ensuremath{\ell^+ \ell^-}\xspace}

\def\kaon  {\ensuremath{K}\xspace}
\def\Kbar  {\kern 0.2em\overline{\kern -0.2em K}{}\xspace}

\def\Kz    {\ensuremath{K^0}\xspace}
\def\Kzb   {\ensuremath{\Kbar^0}\xspace}
\def\KzKzb {\ensuremath{\Kz \kern -0.16em \Kzb}\xspace}
\def\Kp    {\ensuremath{K^+}\xspace}
\def\Km    {\ensuremath{K^-}\xspace}

\def\KpKm  {\ensuremath{\Kp \kern -0.16em \Km}\xspace}

\def\Kstarz  {\ensuremath{K^{*0}}\xspace}

\def\Kstar   {\ensuremath{K^*}\xspace}

\def\Kstarp  {\ensuremath{K^{*+}}\xspace}

\def\Dbar    {\kern 0.2em\overline{\kern -0.2em D}{}\xspace}

\def\Dz      {\ensuremath{D^0}\xspace}
\def\Dzb     {\ensuremath{\Dbar^0}\xspace}
\def\DzDzb   {\ensuremath{\Dz {\kern -0.16em \Dzb}}\xspace}
\def\Dp      {\ensuremath{D^+}\xspace}
\def\Dm      {\ensuremath{D^-}\xspace}

\def\DpDm    {\ensuremath{\Dp {\kern -0.16em \Dm}}\xspace}

\def\B       {\ensuremath{B}\xspace}
\def\Bbar    {\kern 0.18em\overline{\kern -0.18em B}{}\xspace}

\def\BB      {\ensuremath{B\Bbar}\xspace} 
\def\Bz      {\ensuremath{B^0}\xspace}
\def\Bzb     {\ensuremath{\Bbar^0}\xspace}
\def\BzBzb   {\ensuremath{\Bz {\kern -0.16em \Bzb}}\xspace}
\def\Bu      {\ensuremath{B^+}\xspace}
\def\Bub     {\ensuremath{B^-}\xspace}

\def\BpBm    {\ensuremath{\Bu {\kern -0.16em \Bub}}\xspace}

\def\BorBbar    {\kern 0.18em\optbar{\kern -0.18em B}{}\xspace}
\def\DorDbar    {\kern 0.18em\optbar{\kern -0.18em D}{}\xspace}
\def\KorKbar    {\kern 0.18em\optbar{\kern -0.18em K}{}\xspace}

\def\psitwos  {\ensuremath{\psi{(2S)}}\xspace}

\mathchardef\Upsilon="7107
\def\Y#1S{\ensuremath{\Upsilon{(#1S)}}\xspace}% no space before {...}!

\mathchardef\Deltares="7101
\mathchardef\Xi="7104
\mathchardef\Lambda="7103
\mathchardef\Sigma="7106
\mathchardef\Omega="710A

\def\Deltabar{\kern 0.25em\overline{\kern -0.25em \Deltares}{}\xspace}
\def\Lbar{\kern 0.2em\overline{\kern -0.2em\Lambda\kern 0.05em}\kern-0.05em{}\xspace}
\def\Sigbar{\kern 0.2em\overline{\kern -0.2em \Sigma}{}\xspace}
\def\Xibar{\kern 0.2em\overline{\kern -0.2em \Xi}{}\xspace}
\def\Obar{\kern 0.2em\overline{\kern -0.2em \Omega}{}\xspace}
\def\Nbar{\kern 0.2em\overline{\kern -0.2em N}{}\xspace}
\def\Xb{\kern 0.2em\overline{\kern -0.2em X}{}\xspace}

\def\BR         {{\ensuremath{\cal B}\xspace}}

\def\mes        {\mbox{$m_{\rm ES}$}\xspace}

\def\DeltaE     {\mbox{$\Delta E$}\xspace}

\newcommand{\tev}{\ensuremath{\mathrm{\,Te\kern -0.1em V}}\xspace}
\newcommand{\gev}{\ensuremath{\mathrm{\,Ge\kern -0.1em V}}\xspace}
\newcommand{\mev}{\ensuremath{\mathrm{\,Me\kern -0.1em V}}\xspace}
\newcommand{\kev}{\ensuremath{\mathrm{\,ke\kern -0.1em V}}\xspace}
\newcommand{\ev}{\ensuremath{\mathrm{\,e\kern -0.1em V}}\xspace}
\newcommand{\gevc}{\ensuremath{{\mathrm{\,Ge\kern -0.1em V\!/}c}}\xspace}
\newcommand{\mevc}{\ensuremath{{\mathrm{\,Me\kern -0.1em V\!/}c}}\xspace}
\newcommand{\gevcc}{\ensuremath{{\mathrm{\,Ge\kern -0.1em V\!/}c^2}}\xspace}
\newcommand{\mevcc}{\ensuremath{{\mathrm{\,Me\kern -0.1em V\!/}c^2}}\xspace}
\def\invfb   {\ensuremath{\mbox{\,fb}^{-1}}\xspace}

\def\mus  {\ensuremath{\rm \,\mus}\xspace}

\def\mus        {\ensuremath{\,\mu{\rm s}}\xspace}    %% microsecond
\def\to                 {\ensuremath{\rightarrow}\xspace}

\def\pep2{PEP-II}

\def\gsim{{~\raise.15em\hbox{$>$}\kern-.85em
          \lower.35em\hbox{$\sim$}~}\xspace}
\def\lsim{{~\raise.15em\hbox{$<$}\kern-.85em
          \lower.35em\hbox{$\sim$}~}\xspace}

\xspace

\def\jetset74   {\mbox{\tt Jetset \hspace{-0.5em}7.\hspace{-0.2em}4}\xspace}

\long\def\ignore#1{\relax}

\newcommand{\BABARPubYear}    {03}
\newcommand{\BABARPubNumber}  {021}
          
\newcommand{\SLACPubNumber} {10132}

\newcommand{\tbllabel}[1]{\textbf{#1}}

\def\D {\ensuremath{D}\xspace}
\def\K {\ensuremath{K}\xspace}

\def\Kmaybestar {\ensuremath{K^{(*)}\xspace}}

\def\kll {\B\to\Kmaybestar\ellell\xspace}

\def\mkpi {\ensuremath{m_{\kaon\pi}}\xspace}

\def\figurebox#1#2#3{%
    \def\arg{#3}%
    \ifx\arg\empty
    {\hfill\vbox{\hsize#2\hrule\hbox to #2{\vrule\hfill\vbox to #1{\hsize#2\vfill}\vrule}\hrule}\hfill}%
    \else
    {\hfill\epsfbox{#3}\hfill}%
    \fi}
 
\long\def\inst#1{\par\nobreak\kern 4pt\nobreak
    {\it #1}\par\vskip 10pt plus 3pt minus 3pt}

\begin{document}

\preprint{\babar-PUB-\BABARPubYear/\BABARPubNumber}
\preprint{SLAC-PUB-\SLACPubNumber}
 
\begin{flushleft}
%\babar\ Analysis Document \# 632, Version 12\\   
\babar-PUB-\BABARPubYear/\BABARPubNumber\\
SLAC-PUB-\SLACPubNumber\\ 
%hep-ex/\LANLNumber\\ [10mm]
\end{flushleft}
\title{
%\vskip 10mm
{\large \bf \boldmath Evidence for the Rare Decay 
$B \rightarrow K^*\ell^+ \ell^-$
and Measurement of the $B \rightarrow K\ell^+ \ell^-$ Branching Fraction}
}
%\begin{center} % Temporary fix for Collaboration name problem
%\vskip 10mm
%The \babar\ Collaboration
%\end{center}
% Input author list
%\input pubboard/authors_jul2003.tex         
%% author list as of 02-Jul-2003 (593 authors)
%
\author{B.~Aubert}
\author{R.~Barate}
\author{D.~Boutigny}
\author{J.-M.~Gaillard}
\author{A.~Hicheur}
\author{Y.~Karyotakis}
\author{J.~P.~Lees}
\author{P.~Robbe}
\author{V.~Tisserand}
\author{A.~Zghiche}
\affiliation{Laboratoire de Physique des Particules, F-74941 Annecy-le-Vieux, France }
% ...
\collaboration{The \babar\ Collaboration}
\noaffiliation

\date{\today}

% Abstract
\begin{abstract}
We present evidence for the flavor-changing neutral current decay
$B\to K^*\ell^+\ell^-$ and a measurement of the branching fraction for
the related process $B\to K\ell^+\ell^-$, where $\ell^+\ell^-$ is
either an $e^+e^-$ or $\mu^+\mu^-$ pair.  These decays are highly
suppressed in the Standard Model, and they are sensitive to
contributions from new particles in the intermediate state.  The data
sample comprises $123\times 10^6$ $\Upsilon(4S)\to \BB$ decays
collected with the \babar\ detector at the \pep2 $e^+e^-$ storage
ring.  Averaging over \Kmaybestar\ isospin and lepton flavor, we
obtain the branching fractions ${\mathcal B}(B\to
K\ell^+\ell^-)=(0.65^{+0.14}_{-0.13}\pm 0.04)\times 10^{-6}$ and
${\mathcal B}(B\to K^*\ell^+\ell^-)=(0.88^{+0.33}_{-0.29}\pm
0.10)\times 10^{-6}$, where the uncertainties are statistical and
systematic, respectively.  The significance of the $B\to
K\ell^+\ell^-$ signal is over $8\sigma$, while for $B\to
K^*\ell^+\ell^-$ it is $3.3\sigma$.

\end{abstract}
 
% PACS, the Physics and Astronomy Classification Scheme.
\maketitle
\par    
   
% The body of the paper starts here

% introduction

Rare decays of $B$ mesons that involve loop diagrams 
in the Standard Model (SM)
provide a promising means to search for 
effects beyond the SM~\cite{bib:TheoryA}.
The decays $B\to K\ell^+\ell^-$ and $B\to K^*\ell^+\ell^-$,
where $\ell^{\pm}$ are charged leptons and $K^*$ is the $K^*(892)$ meson, 
result from one-loop processes that transform the $b$-quark
in the initial-state $B$ meson into an $s$-quark in the final-state
$K^{(*)}$ meson. 

In the SM, three amplitudes
contribute at leading order: an electromagnetic (EM) penguin,
a $Z$ penguin, and a $W^+W^-$ box diagram. 
The penguin diagrams involve the emission and absorption of a $W$ boson.
The presence
of three SM electroweak amplitudes makes $B\to K^{(*)}\ell^+\ell^-$
more complex than $B\to K^*\gamma$, which proceeds solely through 
an EM penguin.
 
Because of their loop structure, these decays are highly
suppressed, with SM branching fractions expected to be 
roughly $0.5\times 10^{-6}$ for $B\to K\ell^+\ell^-$ and
about three times that for the $B\to K^{*}\ell^+\ell^-$ 
modes~\cite{bib:TheoryA,bib:TheoryB,bib:TheoryC}. 
Due to the complexity of strong interaction effects, however, 
theoretical
uncertainties on the rates are currently at least 35\% 
(Ali \textit{et al.}~\cite{bib:TheoryA}).   
Both $B\to K^{*}e^+e^-$ and $B\to K^*\mu^+\mu^-$
receive a contribution from the pole in
the EM penguin amplitude at $q^2=m_{\ell^+\ell^-}^2=0,$
but the enhancement in the electron mode is larger.
An important consequence of the loop structure of
these decays is that their branching fractions and 
kinematic distributions can be significantly affected by the presence of   
new particles, such as those predicted in models based on
supersymmetry~\cite{bib:TheoryA}. 

Recently, substantial progress has been made 
in experimental studies of these decays. 
The Belle collaboration has observed
$B\to K\ell^+\ell^-$, as well as the inclusive
$B\to X_s\ell^+\ell^-$ decay~\cite{bib:Belle}. 
Limits on these and similar modes 
have been set by \babar ~\cite{bib:BaBarPRL2002}, 
Belle~\cite{bib:Belle}, CLEO~\cite{bib:CLEO}, and
CDF~\cite{bib:CDF}.
Our new measurements are based on a data sample
six times larger than that used for our previously published results. 
We study eight final states:
$B^+\to K^+\ell^+\ell^-$, 
$B^0\to K_S^0\ell^+\ell^-$,
$B^+\to K^{*+}\ell^+\ell^-$, and 
$B^0\to K^{*0}\ell^+\ell^-$, where
$K^{*0}\to K^+\pi^-$, $K^{*+}\to K_S^0\pi^+$,
$K_S^0\to\pi^+\pi^-$, and
$\ell$ is either an 
$e$ or $\mu$.
Throughout this paper, charge-conjugate modes are implied.

%  Data sample

We analyze data collected with the \babar\ detector~\cite{bib:babarNIM}
at the \pep2\ storage ring at the Stanford Linear Accelerator Center. 
The data sample comprises 113.1 \invfb\ recorded on the 
$\Upsilon(4S)$ resonance, yielding $(122.9\pm1.4)\times 10^6$ $\BB$ decays, and
an off-resonance sample of 12.0 \invfb\ used to study 
continuum background.

% event selection

We select events that include two
oppositely charged leptons ($e^+e^-$, $\mu^+\mu^-$), a kaon (either
$K^{\pm}$ or $K_S^0$), and, for the $B\to K^{*}\ell^+\ell^-$ modes, a
$\pi^{\pm}$ that combines with a kaon to form a $K^*$ candidate.
Electrons are identified primarily in the CsI(Tl) electromagnetic calorimeter,
while muons are identified by their penetration through 
iron plates of the magnet flux return.
Electron (muon) candidates are required to satisfy 
$p > 0.5\ (1.0) \ {\rm GeV}/c$. 
Bremsstrahlung photons from electrons are recovered by combining an 
electron candidate with up
to one photon with $E_{\gamma}>30\ {\rm MeV}$ in a small angular
region around the initial electron direction. 
Photon conversions and $\pi^0$ Dalitz decays are removed 
by vetoing all low-mass $e^+e^-$ pairs, except in 
$B\to K^*e^+e^-$ modes, where we preserve acceptance at low mass 
by retaining pairs that intersect
inside the beam pipe.

$K^{\pm}$ candidates are tracks with $dE/dx$ 
and Cherenkov angle consistent
with a kaon.  $\pi^{\pm}$ candidates are tracks that do not satisfy 
the $K^{\pm}$ selection.  $K^0_S$ candidates are
reconstructed from two oppositely charged tracks with an invariant
mass consistent with the $K^0_S$ mass and a common vertex displaced
from the primary vertex by at least 1 mm.  

True $B$ signal decays produce narrow peaks in the distributions of two
kinematic variables, which can be fitted to extract
the signal and background yields.
For a candidate system
of $B$ daughter particles with
masses $m_{i}$ and 
three-momenta ${\bf p}^*_{i}$ in the $\Upsilon(4S)$ center-of-mass (CM) frame,
we define $m_{\rm ES}= \sqrt{E_{\rm b}^{*2} - |\sum_{i} {\bf p}^*_{i}|^2}$
and $\Delta E= \sum_{i}\sqrt{m_i^2 + {\bf p}_{i}^{*2}}- E_{\rm b}^*$,
where $E_{\rm b}^*$ is the beam energy in the CM frame. 
For signal events, the $m_{\rm ES}$ distribution peaks at the $B$ meson 
mass with resolution $\sigma\approx 2.5\ {\rm MeV}/c^2$, 
and the $\Delta E$
distribution peaks near zero, with a typical width $\sigma \approx$ 20 MeV.
In $B\to K\ell^+\ell^-$ channels, we perform
a two-dimensional unbinned maximum-likelihood fit to the
distribution of $m_{\rm ES}$ and $\Delta E$ in the region 
$m_{\rm ES}>5.2\ {\rm GeV}/c^2$
and $|\Delta E|<0.25$ GeV. In $B\to \Kstar\ell^+\ell^-$ decays,
we perform a three-dimensional fit to $m_{\rm ES}$, $\Delta E$,  
and the kaon-pion invariant mass
in the region
$ 0.7 < m_{K\pi} < 1.1\ {\rm GeV}/c^2$.

Backgrounds arise from three main sources: random combinations of
particles from $q\bar q$ events produced in the continuum, random
combinations of particles from $\Upsilon(4S)\to \BB$ decays, and $B$
decays to topologies similar to the signal modes.  The first two
(``combinatorial'') backgrounds typically arise from pairs of
semileptonic decays and produce broad distributions in \mes and
\DeltaE compared to the signal.  The third source arises from modes such as $B\to
J/\psi K^{(*)}$ (with $J/\psi\to\ell^+\ell^-$) or $B\to K^{(*)}\pi\pi$
(with pions misidentified as muons), which have shapes similar to
the signal.
All selection criteria are optimized with \texttt{GEANT}4~\cite{bib:GEANT} 
simulated data or with data samples outside
the full fit region.

We suppress combinatorial background from continuum processes using a
Fisher discriminant~\cite{bib:Fisher}, which is a linear combination of
variables with coefficients optimized to distinguish between signal
and background. The variables (defined in the CM frame) are 
(1) the ratio of second- to zeroth-order Fox-Wolfram
moments~\cite{bib:FoxWolfram} for the event, computed using all
charged tracks and neutral energy clusters;
(2) the angle between the thrust axis of the $B$ candidate and that of
the remaining particles in the event;
(3) the production angle $\theta_B$ of the $B$ candidate with respect
to the beam axis; and
(4) the masses of $K\ell$ pairs with charge correlation
 consistent with $D$ decay. 

We suppress combinatorial backgrounds from $\BB$ events
using a likelihood function constructed
from (1) the missing energy
of the event, computed from all charged tracks and 
neutral energy clusters; (2) the vertex
fit probability of all tracks from the $B$ candidate;
(3) the vertex fit probability of the two leptons; and
(4) the angle $\theta_B$.
Missing energy provides the strongest suppression of
combinatorial $B\bar B$ background events, which typically 
contain neutrinos from two semileptonic decays.

The most prominent backgrounds that peak in $m_{ES}$ and $\Delta E$
are $B$ decays to charmonium: $B\to J/\psi K^{(*)}$
(with $J/\psi \to\ell^+\ell^-$) and analogous $B$ decays to $\psitwos$. 
We exclude dilepton pairs consistent with the $J/\psi$
($2.90<m_{e^+e^-}<3.20\ {\rm GeV}/c^2$ and $3.00<m_{\mu^+\mu^-}<3.20\ {\rm GeV}/c^2$)
or with the $\psi(2S)$
($3.60<m_{\ell^+\ell^-}<3.75\ {\rm GeV}/c^2$). 
This veto is also applied to $m_{e^+e^-}$
computed without bremsstrahlung photon recovery.
When a
lepton radiates or is mismeasured, $m_{\ell^+\ell^-}$ can shift away
from the charmonium mass, while $\Delta E$ shifts in a correlated
manner. The veto region is extended in the $(m_{\ell^+\ell^-},\Delta E)$ plane
to account for this correlation,
removing nearly all charmonium events and
simplifying the description of the background in the fit.  Because the
charmonium events removed by these vetoes are so similar to signal
events, these modes provide extensive control samples (about 5200
events in all) for studying signal shapes, selection efficiencies, and
systematic errors.  Outside the charmonium veto regions, the signal
efficiency is similar over the full $q^2$ range of each mode.

In muon modes, where the probability for a hadron to be
misidentified as a muon can be as high as a few percent,
background from the decay $B^-\to D^0 \pi^-$ with
$D^0\to K^-\pi^+$ or $D^0\to K^{*-}\pi^+$,
or from $\bar B^0\to D^+\pi^-$ with $D^+\to \bar K^{*0}\pi^+$, 
is significant.  These events are suppressed by vetoing events where
the $\Kmaybestar\mu$ kinematics are consistent with those of a
hadronic \D decay.

We estimate the residual peaking background 
from measurements in the 
data, supplemented in some cases by
simulation studies. Events from $B\to K^{(*)}\pi\pi$,
$B\to K^{(*)}K\pi$, and $B\to K^{(*)}KK$ are highly
suppressed by the particle identification criteria. These
backgrounds are estimated from control samples 
to be $0.19\pm0.11$ events per channel averaged over muon modes and  
less than $0.01$ events per channel in electron modes. 
After the vetoes on $B\to J/\psi K^{(*)}$ and $B\to \psitwos K^{(*)}$
decays, the remaining peaking background is estimated from simulation
to be $0.17\pm0.07$ events per channel averaged over 
$B\to K^*\ell^+\ell^-$ modes, and it is
negligible in $B\to K\ell^+\ell^-$ modes.  
The background from $B\to K^*\gamma$ (with
photon conversion in the detector) is determined from simulation to be
$0.48\pm0.16$ events in $B^0\to K^{*0}e^+e^-$ and $0.09\pm 0.04$
events in $B^+\to K^{*+}e^+e^-$.

\begin{table}
 \caption[Fit results]{ Results from the fits to \kll modes.  The
  columns are, from left to right: fitted signal yield;
  the signal efficiency,
  $\epsilon$ (not including the branching fractions 
  for \Kstar and \Kz decays); the fractional systematic error on the selection efficiency,
  $\Delta\BR_{\epsilon}/\BR_{\epsilon}$; the systematic error from the fit,
  $\Delta\BR_{\rm fit}$; and the branching fraction central value
  (\BR) with its statistical and total systematic
  uncertainties. 
  For the branching fractions averaged over different channels 
  (lower part of table), simultaneous, constrained fits are 
  performed to extract an efficiency-corrected signal yield that 
  averages over the included channels.  The modes with significance 
  $> 3\sigma$ are $\Kp\epem$ ($8.4\sigma$), $\Kz\mumu$ ($4.1\sigma$), $\K\epem$ ($7.8\sigma$), $\K\ellell$ ($8.4\sigma$), and $\Kstar\ellell$ ($3.3\sigma$).

}
 \footnotesize
 \begin{center}
 \begin{tabular}{l D{.}{.}{3.5} D{.}{.}{2.1} D{.}{.}{3.1} D{.}{.}{2.1} D{.}{.}{2.11}}
\hline\hline \vspace*{-.6cm} \\
 \multicolumn{1}{c}{\tbllabel{Mode}}
 & \multicolumn{1}{c}{\tbllabel{\parbox[b]{1.0cm}{\center Signal yield}}}
 & \multicolumn{1}{c}{\tbllabel{\parbox[b]{0.8cm}{\center $\epsilon$ \\ (\%)}}}
 & \multicolumn{1}{c}{\tbllabel{\parbox[b]{1.1cm}{\center 
     $\Delta \BR_{\epsilon}/\BR$ (\%)}}}
 & \multicolumn{1}{c}{\tbllabel{\parbox[b]{1cm}{\center
     $\Delta \BR_{\rm fit}$ ($10^{-6}$)}}}
 & \multicolumn{1}{c}{\tbllabel{\parbox[b]{2.4cm}{\center $\BR$ \\ ($10^{-6}$)}}}

 \\ \vspace{-.2cm} \\
 \hline \vspace{-.2cm} \\
$\Kp\epem$             & 24.7^{+5.9}_{-5.2}    &   19.2   &   \pm 6.3   & \pm 0.02  &  1.05^{+0.25}_{-0.22}\pm0.07  \\    
$\Kp\mumu$            &  0.7^{+2.0}_{-1.2}    &   8.5    &   \pm 7.6   & \pm 0.02  &  0.07^{+0.19}_{-0.11}\pm0.02  \\    
$\Kz\epem$             & -1.8^{+2.0}_{-1.4}    &   20.1   &   \pm 8.4   & \pm 0.08  & -0.21^{+0.23}_{-0.16}\pm0.08  \\    
$\Kz\mumu$             &  5.9^{+3.0}_{-2.3}    &    8.6   &  \pm  8.8   & \pm 0.02  &  1.63^{+0.82}_{-0.63}\pm0.14  \\    
$\Kstarz\epem$         & 12.4^{+6.3}_{-5.2}    &   13.6   &   \pm 7.6   & \pm 0.08  &  1.11^{+0.56}_{-0.47}\pm0.11  \\    
$\Kstarz\mumu$    &  4.5^{+4.1}_{-3.0}    &    6.4   &  \pm 10.1   & \pm 0.07  &  0.86^{+0.79}_{-0.58}\pm0.11  \\    
$\Kstarp\epem$        &  0.6^{+3.8}_{-2.5}    &   10.2   &  \pm 10.7   & \pm 0.28  &  0.20^{+1.34}_{-0.87}\pm0.28  \\    
$\Kstarp\mumu$    &  4.2^{+3.5}_{-2.4}    &    4.8   &  \pm 12.7   & \pm 0.15  &  3.07^{+2.58}_{-1.78}\pm0.42  \\ 
\vspace{-.2cm} \\
 \hline \vspace{-.2cm} \\
$\K\epem$       & \multicolumn{1}{r}{$91^{+22}_{-19}$}   & &  \pm 6.5   &  \pm 0.02    &  0.74^{+0.18}_{-0.16}\pm 0.05  \\    
$\K\mumu$       & \multicolumn{1}{r}{$55^{+29}_{-23}$}   & &  \pm 7.4   &  \pm 0.01    &  0.45^{+0.23}_{-0.19}\pm 0.04 \\    
$\K\ellell$     & \multicolumn{1}{r}{$80^{+17}_{-15}$}   & &  \pm 6.4   &  \pm 0.01    &  0.65^{+0.14}_{-0.13}\pm 0.04 \\    
$\Kstar\epem$   & \multicolumn{1}{r}{$121^{+61}_{-51}$}   & &  \pm 7.8   &  \pm 0.08    &  0.98^{+0.50}_{-0.42}\pm 0.11 \\    
$\Kstar\mumu$   & \multicolumn{1}{r}{$156^{+94}_{-75}$}   & &  \pm 10.1  &  \pm 0.09    &  1.27^{+0.76}_{-0.61}\pm 0.16  \\    
$\Kstar\ellell$ & \multicolumn{1}{r}{$108^{+41}_{-36}$}   & &  \pm 8.1   &  \pm 0.07    &  0.88^{+0.33}_{-0.29}\pm 0.10 \\ 
\vspace{-.2cm} \\   
\hline\hline   
\end{tabular}
\end{center}
\label{tab:results}
\end{table}   

The signal shapes are parameterized with a Gaussian core for $m_{\rm
ES}$ and a double Gaussian core for $\Delta E$.  Both the 
$m_{\rm ES}$ and $\Delta E$ shapes include a radiative tail, which 
accounts for the effects of bremsstrahlung.
The $m_{\rm ES}$ shape parameters are assumed to have
$\Delta E$ dependence $c_0 + c_2(\Delta E)^2$.
All signal shape parameters are fixed from 
signal simulation, except for the mean and width parameters
in $m_{\rm ES}$ ($c_0$ only) and $\Delta E$, which are fixed to values 
from charmonium data control samples. 

The background is modeled as the sum of three terms: (1) a
combinatorial background shape with floating
normalization, written as the product
of an ARGUS function~\cite{bib:ARGUS} in $m_{\rm ES}$, an exponential 
in $\Delta E$, and the product of $\sqrt{m_{K\pi}-m_K-m_{\pi}}$ and a 
quadratic function of $m_{K\pi}$ for the 
$K^*$ modes; (2) a peaking background contribution, with
the same shape as the signal, but with 
normalization fixed to measured peaking backgrounds; and
(3) terms with floating normalization 
to describe (a) background in $B\to K\ell^+\ell^-$ ($B\to K^{*}\ell^+\ell^-$)
from $B\to K^*\ell^+\ell^-$ ($B\to K^*\pi\ell^+\ell^-$) 
events with a lost pion, and (b)
background in $B\to K^*\ell^+\ell^-$ from $B\to K\ell^+\ell^-$ events
with a randomly added pion.
In the $K^*$ modes, we allow an additional
background (4) that uses our combinatorial shape in $m_{\rm ES}$
and $\Delta E$, but peaks in $m_{K\pi}$ at the $K^*$ mass.
Because the normalizations for terms (1), (3), and (4) are
floating, as are the combinatorial background shape 
parameters, much of the uncertainty in the background
is propagated into the statistical 
uncertainty on the signal yield obtained from the
fit.

% Systematic errors 

Table~\ref{tab:results} lists 
signal yields and branching fractions for each mode.
The relative systematic uncertainties on the efficiency, 
$\Delta \BR_{\epsilon}/\BR$,
arise from  
charged-particle tracking (1.0\% per lepton, 
1.7\% per charged hadron), 
particle identification (1.1\% per electron,
1.6\% per muon, 0.9\% per pion, 0.9\% per kaon), 
the continuum suppression cut [(0.8--2.8)\%],
the $\BB$ suppression cut [(1.4--5.0)\%], 
$K_S^0$ selection (3.8\%),
signal simulation statistics [(0.7--1.4)\%], 
theoretical model dependence of the efficiency [(4--7)\%, depending
on the mode], and
the number of $\BB$ events (1.1\%).
Uncertainties on efficiencies due to model dependence of 
form factors are taken to be the full range of variation
from a set of models~\cite{bib:TheoryB}.

The systematic uncertainties on the fit yields, 
$\Delta \BR_{\rm fit},$ arise from three sources:
uncertainties in the parameters describing
the signal shapes,
possible correlation between
$m_{\rm ES}$ and $\Delta E$ in
the combinatorial background shape, and uncertainties in
the peaking backgrounds. 
The uncertainties in the means and widths of the 
signal shapes are obtained by comparing data and 
simulation for the charmonium control samples.
For modes with electrons, 
we also vary the fraction
of signal events in the tail of the $\Delta E$ distribution. 
To evaluate sensitivity to the background parameterization,
we allow additional
parameters and a correlation between
$m_{\rm ES}$ and $\Delta E$. 

\begin{figure}[!tbh]
 \begin{center}
   \includegraphics[width=1.0\linewidth]{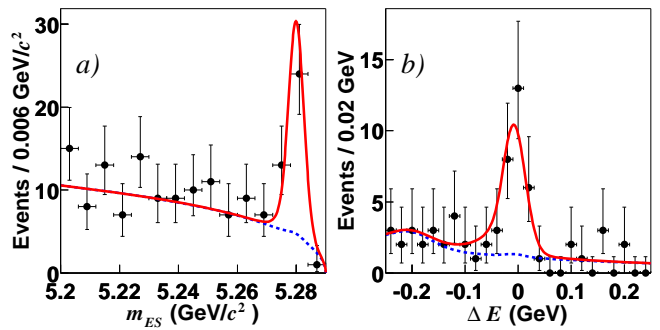}
  \end{center}
  \vspace{-0.5cm}
\caption[kll_combined.]
{\label{fig:kll_combined}
Distributions of the fit variables in $K\ell^+\ell^-$ data (points),
compared with projections of the simultaneous fit (curves): (a) $m_{\rm
ES}$ distribution after requiring $-0.11<\Delta E<0.05\ {\rm GeV}$ and
(b) $\Delta E$ distribution after requiring 
$|m_{\rm ES} - m_{B}| < 6.6 \mevcc$ $(2.6\sigma)$.
The solid curve is the sum of all fit components,
including signal; the dashed curve is the sum of all background
components.
}
\end{figure}   

\begin{figure}[!tbh]
 \begin{center}
   \includegraphics[width=1.0\linewidth]{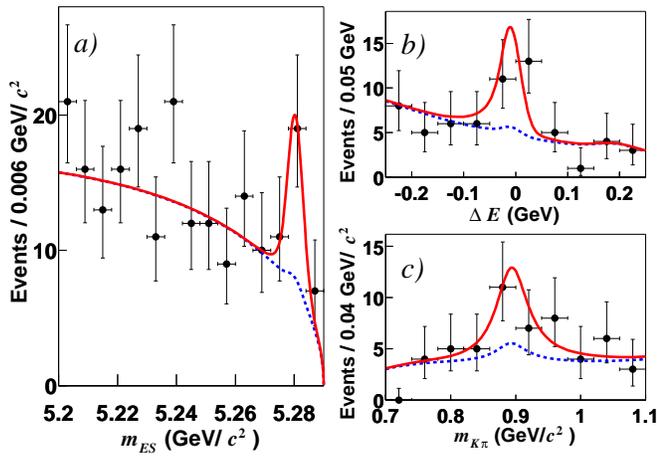}
  \end{center}
  \vspace{-0.5cm}
\caption[kstll_combined.]
{\label{fig:kstll_combined}
Distributions of the fit variables in $K^*\ell^+\ell^-$ data (points),
compared with projections of the simultaneous fit (curves):
(a) \mes after requiring $-0.11<\Delta E<0.05\ {\rm GeV}$
and $0.817< \mkpi <0.967\ {\rm GeV}/c^2$,
(b) $\Delta E$ after requiring 
$|m_{\rm ES} - m_{B}| < 6.6 \mevcc$ $(2.6\sigma)$,
$0.817< \mkpi <0.967\ {\rm GeV}/c^2$, and   
(c) \mkpi after requiring 
$|m_{\rm ES} - m_{B}| < 6.6 \mevcc$
and $-0.11<\Delta E<0.05\ {\rm GeV}$. 
The solid curve is the sum of all fit components, including signal; the 
dashed curve is the sum of all background components.
}
\end{figure}

Table~\ref{tab:results} also lists results from simultaneous
fits to combinations of $B\to K\ell^+\ell^-$ modes and  
combinations of $B\to K^*\ell^+\ell^-$ modes, where the
relative branching fractions for the contributing modes are constrained.
$B^0$ and $B^+$ production rates are constrained to be equal, and the
ratio of their total widths is constrained to be $1.085\pm0.017$~\cite{bib:b life}.
All branching fractions from simultaneous fits are expressed in terms of the
$B^0$ total width.
The projections of the fit on $m_{\rm ES}$ and $\Delta E$ are 
shown in Fig.~\ref{fig:kll_combined} for the
simultaneous fit to the 
$B\to K\ell^+\ell^-$ channels. We assume that all four $B\to K\ell^+\ell^-$ modes
have equal partial widths.
A signal is evident at the $B$
mass in $m_{\rm ES}$ and at $\DeltaE=0$.
Figure~\ref{fig:kstll_combined} shows projections of the simultaneous
fit to all $B\to K^*\ell^+\ell^-$ modes.
Here, the partial width ratio of electron and muon modes is constrained to
be $\Gamma(B\to K^*e^+e^-)/\Gamma(B\to K^*\mu^+\mu^-)=1.33$ from
the model of Ali {\it et al.}~\cite{bib:TheoryA}.  Our 
simultaneous fit result is expressed as a
$\Bz\to \Kstarz\mumu$ branching fraction.  

The significance of the $B\to K\ell^+\ell^-$ signal from the
simultaneous fit is 
$\sim 8\sigma$, computed as $\sqrt{2\Delta\log{\cal L}}$, where
$\Delta\log{\cal L}$ is the likelihood difference between the best fit
and the null-signal hypothesis.  We account for systematic
uncertainties in the significance by simultaneously including all
effects that individually lower the fit yields prior to computing the
change in likelihood.  The significance of the $B\to K^*\ell^+\ell^-$
signal, including all systematic uncertainties, is $3.3\sigma$
($3.8\sigma$ not including them).

%  Conclusion section
In summary, 
we have observed signals for $B\to K\ell^+\ell^-$, averaged over
lepton type ($e^+e^-$ and $\mu^+\mu^-$)
and $B$ charge, and we have obtained the first evidence for $B\to K^*\ell^+\ell^-$,
similarly averaged. We obtain
\begin{eqnarray}
{\mathcal B}(B\to K\ell^+\ell^-)=(0.65^{+0.14}_{-0.13}\pm 0.04)\times 10^{-6}, \nonumber\\ 
{\mathcal B}(B\to K^*\ell^+\ell^-)=(0.88^{+0.33}_{-0.29}\pm 0.10)\times 10^{-6},\nonumber 
\end{eqnarray}
where the first error is statistical and the second is systematic.
Our branching fraction for $B \to K\ell^+\ell^-$ is slightly higher than our previous limit $0.51\times10^{-6}$ (90\% confidence level)~\cite{bib:BaBarPRL2002} and is in agreement with the Belle result $(0.75^{+0.25}_{-0.21}\pm 0.09)\times10^{-6}$~\cite{bib:Belle}. Our $B \to K^*\ell^+\ell^-$ branching fraction is consistent with previous 90\% confidence level limits from \babar\ ($<3.1\times 10^{-6}$ for $K^*\ell^+\ell^-$) ~\cite{bib:BaBarPRL2002} and Belle 
($<3.1\times 10^{-6}$ for $K^*\mu^+\mu^-$)~\cite{bib:Belle}.
These results are consistent with the range of predictions based on
the Standard Model~\cite{bib:TheoryA,bib:TheoryB,bib:TheoryC}.

%%%%%%%%%%%%%%%%%%%%%%%%%%%%%%%%%%%%%%%%%%%%%%%%%%%%%%%%%%%%%%%%%%%%%%%%

% Standard acknowledgments paragraph; must always be included.
%\input pubboard/acknow_PRL
We are grateful for the excellent luminosity and machine conditions
provided by our \pep2\ colleagues, 
and for the substantial dedicated effort from
the computing organizations that support \babar.
The collaborating institutions wish to thank 
SLAC for its support and kind hospitality. 
This work is supported by
DOE
and NSF (USA),
NSERC (Canada),
IHEP (China),
CEA and
CNRS-IN2P3
(France),
BMBF and DFG
(Germany),
INFN (Italy),
FOM (The Netherlands),
NFR (Norway),
MIST (Russia), and
PPARC (United Kingdom). 
Individuals have received support from the 
A.~P.~Sloan Foundation, 
Research Corporation,
and Alexander von Humboldt Foundation.

%%%%%%%%%%%%%%%%%%%%%%%%%%%%%%%%%%%%%%%%%%%%%%%%%%%%%%%%%%%%%%%%%%%%%%%%

\end{document}